\documentclass[11pt,twoside]{article}
\usepackage{asp2014}

\aspSuppressVolSlug
\resetcounters

\begin{document}

\title{RTApipe, a framework to develop astronomical pipelines for the real-time analysis of scientific data.}

\author{N.~Parmiggiani,$^{1,2}$ A.~Bulgarelli,$^1$ D.~Beneventano,$^2$, V.~Fioretti,$^1$ L.~Baroncelli,$^1$ A.~Addis,$^1$ and M.~Tavani$^3$}
\affil{$^1$INAF OAS Bologna, Via P. Gobetti 93/3, 40129 Bologna, Italy. \email{nicolo.parmiggiani@inaf.it}}

\affil{$^2$Universit\`{a} degli Studi di Modena e Reggio Emilia, DIEF - Via Pietro Vivarelli 10, 41125 Modena, Italy. }

\affil{$^3$INAF-IAPS Roma, via del Fosso del Cavaliere 100, I-00133 Roma, Italy.}

\paperauthor{Nicol\`{o}~Parmiggiani}{nicolo.parmiggiani@inaf.it}{0000-0002-4535-5329}{INAF}{OAS}{Bologna}{BO}{40129}{Italy}
\paperauthor{Andrea~Bulgarelli}{andrea.bulgarelli@inaf.it}{0000-0001-6347-0649}{INAF}{OAS}{Bologna}{BO}{40129}{Italy}
\paperauthor{Domenico~Beneventano}{domenico.beneventano@unimore.it}{0000-0001-6616-1753}{UNIMORE}{DIEF}{Modena}{MO}{41125}{Italy}
\paperauthor{Valentina~Fioretti}{valentina.fioretti@inaf.it}{0000-0002-6082-5384}{INAF}{OAS}{Bologna}{BO}{40129}{Italy}
\paperauthor{Leonardo~Baroncelli}{leonardo.baroncelli@inaf.it}{0000-0002-9215-4992}{INAF}{OAS}{Bologna}{BO}{40129}{Italy}
\paperauthor{Antonio~Addis}{antonio.addis@inaf.it}{0000-0002-0886-8045}{INAF}{OAS}{Bologna}{BO}{40129}{Italy}
\paperauthor{Marco~Tavani}{marco.tavani@inaf.it}{0000-0003-2893-1459}{INAF}{IAPS}{Roma}{RO}{00133}{Italy}




\begin{abstract}

In the multi-messenger era, astronomical projects share information about transients phenomena issuing science alerts to the Scientific Community through different communications networks. This coordination is mandatory to understand the nature of these physical phenomena. For this reason, astrophysical projects rely on real-time analysis software pipelines to identify as soon as possible transients (e.g. GRBs), and to speed up external alerts' reaction time. These pipelines can share and receive the science alerts through the Gamma-ray Coordinates Network.
This work presents a framework designed to simplify the development of real-time scientific analysis pipelines. The framework provides the architecture and the required automatisms to develop a real-time analysis pipeline, allowing the researchers to focus more on the scientific aspects.  The framework has been successfully used to develop real-time pipelines for the scientific analysis of the AGILE space mission data. It is planned to reuse this framework for the Super-GRAWITA and AFISS projects. A possible future use for the Cherenkov Telescope Array (CTA) project is under evaluation.
  
\end{abstract}

\section{Introduction}

In recent years, the astrophysical landscape changed, and now the so-called "multi-messenger era" is leading part of the observatories activities. In this context, the astrophysical projects share their information with the community to study the same transient event with different "messenger" signals: electromagnetic radiation, gravitational waves, and neutrinos.

Thanks to networks like the Gamma-Ray Coordinates Network (GCN), the coordination of different projects is possible. The framework presented in this work focus on the GCN used by observatories to share science alerts. A science alert is a communication from/to the astrophysical community that a transient phenomenon occurs in the sky. In this multi-messenger context, the observatories need fast analysis to identify transient phenomena and share them with the community. 

We designed and developed a framework to realise flexible Real-Time Scientific Analysis pipelines to fulfil gamma-ray observatories' requirements in this multi-messenger era. The solutions adopted in this framework derive from the know-how acquired during the development of pipelines for the AGILE space mission \citep{2014ApJ...781...19B,2019ExA....48..199B,2019RLSFN.tmp...30B}.

\section{Framework Architecture}

The architecture of this framework is designed to fulfil all the features requested in the multi-messenger context. The RTApipe framework is composed of the following components: Data Model, Science Logic, Pipeline Manager, and Task Manager. This framework can be configured with external components: Data Sources, Science Tools, Analysis Results, and External Interfaces.

The Data Model (DM) consists of all the entities present in this framework: Data Sources, Instruments, Observations, Analysis, Science Tools, and more. The DM is very flexible and allows the RTApipe framework to implement real-time scientific analysis pipelines for different project types such as space missions and ground-based telescope arrays. The developers of the pipeline can configure the DM to fit their use cases. A part of this DM is static and must be configured manually by the user before starting the data acquisition and the real-time analysis with the pipeline. The other part is dynamically updated during the operations by the pipelines. 

The Science Logic (SL) defines the set of rules used by the pipeline to know \textit{When} and \textit{How} execute the configured analysis during operations. This component interacts with most of the dynamic part of the DM. It manages pipeline behaviour during the operations without requiring human intervention. This set of rules, defined in advance, allows the pipeline to operate in real-time autonomously. We developed the framework to perform new analyses when one of the following two conditions occurs. The first condition occurs when a data source updates the data index into the database and then communicates to the pipeline that new data has arrived. The second condition occurs when the pipeline receives an External or Internal Science Alert from other missions through the External Interfaces (e.g. GCN). This last condition is used to search for a counterpart of the Science Alert inside the data archive. The pipeline verifies, using the data index if the data archive contains the required time window for the analysis. If yes, the analysis is performed; otherwise, the pipeline waits until the required data arrives to start suspended analyses.

The Pipeline Manager (PM) and the Task Manager (TM) execute the analysis and manage priority between different tasks and queues. The PM component executes the analyses generated by the SL rules on the available computational resources. The PM can manage more pipelines (e.g. one for each different instrument on a satellite).  Since the number of these analyses can be potentially very high, the pipeline manager submits these processes to a TM component, which executes them in parallel, optimising the computational resources. The TM performs analyses on multiple computing nodes and coordinates them.

\begin{figure*}
	\centering
	  \includegraphics[width=\textwidth]{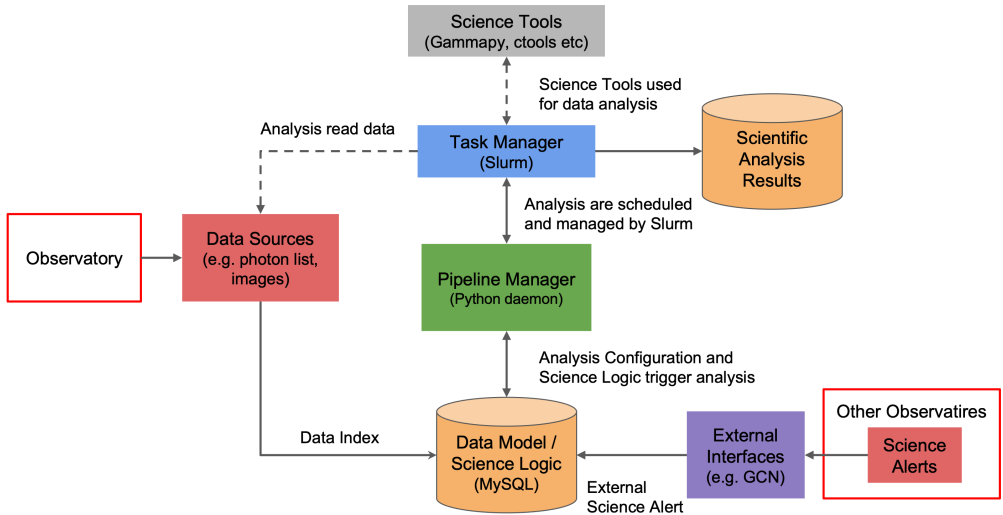}
	\caption{High-level schema of the RTApipe Architecture}
	\label{fig:architecture}
\end{figure*}

The Data Source (DS) is a system that generates one or more Data Types. These data are the input for the analysis of the scientific pipeline. The pipelines developed with this framework can be interfaced with one or more DS. The RTApipe framework does not specify where data should be stored but requires an updated index for each DS. A good example of a DS is the event list generated from an instrument and stored in a database. In the case of a telescope, the event list is the list of triggers with their trigger time, and the index to be updated could be the trigger time for the last triggered event. This value is updated inside the DM. In this way, the pipeline knows when new data arrives into the system, and it also knows the time window of these data on which it can start the analysis. 

The Science Tools (ST) are software developed to perform Scientific Analysis. The pipeline can manage different ST, and these tools can analyse the same data with different parameters or configurations. The pipeline automatically performs the configured analyses with the required ST. The DM contains templates of the configuration files for each ST enabled to work with the pipeline. During the observation, the pipeline uses the templates to create the right configuration for different analysis types.

The RTApipe framework does not require the direct management of the analysis results. Each ST can save its results in different storage systems: database, file system, cloud services, and more.

\section{Framework key features} \label{sec:key_features}

This RTApipe framework has several key features essential for the multi-messenger context:

\begin{itemize}
  \item Parallel analysis: pipelines developed with the RTApipe framework can perform many parallel analyses using the TM. It is possible to run parallel analyses using the same ST with different configuration parameters. Parallel analyses can have the same input data stream or multiple input data streams. Consider a telescope array divided into sub-arrays, each with a different pointing in the sky. The framework can manage a list of analyses performed in parallel for each sub-array. 
  
  \item Flexibility: pipelines can receive several input data flows, use various types of ST, and all this can be configured in just a few simple steps into the DM. The framework does not require developers to use a specific data format or storage system. The RTApipe allows the user to configure easily new ST when needed.
  
  \item Scalability: Slurm (\url{https://slurm.schedmd.com}) is the job scheduler of this framework. Slurm can be configured in a single machine or on thousands of machines by modifying a few lines in the Slurm configuration file. It is possible to increase the computing power without changing the DM or the SL.
  
  \item Task Priority: it is possible to configure the priority between different types of analysis. This feature is important to start high-priority jobs and suspending low priority jobs (e.g. when a Scientific Alert is received).
  
  \item Pipeline Monitoring: the TM saves all the information related to the analysis's execution into a database. Therefore, it is straightforward to perform queries on the database to obtain real-time monitoring or statistical analysis.
  
  \item Easy deployment: the RTApipe framework is configured within a Singularity container (\url{https://sylabs.io/docs/}). All the services needed by the pipelines are installed and configured inside this container. It is possible to deploy the container on different hardware.
  
\end{itemize}

\section{Conclusion}

The multi-messenger astronomy is leading the activities of many gamma-ray observatories and space missions. These projects rely on real-time analysis software pipelines to identify as soon as possible transients (e.g. GRBs) and share the science alerts with the scientific community through networks (e.g, GCN).
This work presents a framework designed to help the development of real-time scientific analysis pipelines. The RTApipe framework allows the researchers to focus more on the software's scientific aspects than on the architecture.  
This framework was used to develop several AGILE real-time analysis pipelines. These pipelines are designed to have a quick response to external science alerts (Gamma-Ray Burst or Gravitational Waves) and analyse the data received from multiple instruments onboard the AGILE satellite. It is planned to reuse this framework for the Super-GRAWITA and AFISS projects. A possible future use for the Cherenkov Telescope Array (CTA) project is under evaluation.

\acknowledgements The AGILE Mission is funded by the Italian Space Agency (ASI) with scientific and programmatic participation by the Italian Institute
of Astrophysics (INAF) and the Italian Institute of Nuclear
Physics (INFN). Investigation supported by the ASI grant I/089/06/2, I/028/12/5 and I/028/12/6.
We thank the ASI management for unfailing support during AGILE operations.
We acknowledge the effort of ASI and industry personnel in operating
the  ASI ground station in Malindi (Kenya), and the data processing done
at the ASI/SSDC in Rome: the success of AGILE scientific operations depends on the effectiveness of the data flow from Kenya to SSDC and INAF/OAS Bologna and the data analysis and software management.


\bibliography{P4-24}


\end{document}